\documentclass{article}

\usepackage{arxiv}

\usepackage[utf8]{inputenc} 
\usepackage[T1]{fontenc}    
\usepackage{hyperref}       
\usepackage{booktabs}       
\usepackage{amsfonts}       
\usepackage{nicefrac}       
\usepackage{microtype}      
\usepackage{lipsum}
\usepackage{graphicx}
\graphicspath{ {./images/} }
\usepackage{graphicx}
\usepackage{array}
\usepackage{tabularx}
\usepackage{multirow}
\usepackage{xcolor}
\usepackage{cite}
\usepackage{url}
\usepackage{tikz}
\usetikzlibrary{positioning,arrows.meta,fit,backgrounds,calc,shapes.geometric}

\newcolumntype{L}[1]{>{\raggedright\arraybackslash}p{#1}}
\newcolumntype{Y}{>{\raggedright\arraybackslash}X}

\tikzset{
  box/.style={rectangle, rounded corners=2pt, draw=black!70, line width=0.5pt,
              align=center, inner sep=4pt, font=\footnotesize},
  lane/.style={rectangle, rounded corners=3pt, draw=black!55, line width=0.6pt,
               align=center, inner sep=6pt, font=\small\bfseries},
  cat/.style={rectangle, rounded corners=2pt, draw=black!70, fill=black!6,
              align=center, inner sep=4pt, font=\footnotesize\bfseries},
  leafn/.style={rectangle, rounded corners=1.5pt, draw=black!45, align=left,
               inner sep=3pt, font=\scriptsize},
  flow/.style={-{Stealth[length=2.2mm]}, line width=0.6pt, draw=black!70},
  dflow/.style={-{Stealth[length=2.2mm]}, line width=0.6pt, draw=black!55, dashed},
}

\title{Human-AI Collaboration and the Transformation of Software Engineering Work}

\author{
 Mamdouh Alenezi \\
  Saudi Data and Artificial Intelligence (SDAIA)\\
  Riyadh, Saudi Arabia
}

\begin{document}
\maketitle
\begin{abstract}
The integration of Generative AI (GenAI) and Agentic AI into software development
is reconfiguring software engineering from an activity centered on human authorship
of code into a discipline centered on directing, verifying, and governing
autonomous and semi-autonomous systems. Drawing on a curated, multi-source evidence
base of recent peer-reviewed and archival studies---including large-scale empirical
observations of autonomous coding agents contributing hundreds of thousands of pull
requests to open-source repositories---this paper synthesizes how the locus of
engineering work is shifting from individual coding productivity toward human--AI
collaboration, agent orchestration, verification and validation, governance, and
socio-technical systems thinking. We adopt a structured interpretive synthesis to
characterize three coexisting paradigms: Traditional, Generative AI-Enabled, and
Agentic AI-Enabled software engineering. We map which traditional activities are
being automated, which are being augmented, and which are newly emerging, and we
trace plausible role trajectories over the next decade. The paper's principal
contribution is an original, theory-driven competency framework that organizes the
capabilities required of future engineers into five interacting categories---%
technical, cognitive, socio-technical, governance, and organizational---%
operationalized through a competency matrix and a transformation framework linking
paradigm shifts to capability demands. We derive nine empirically testable
propositions and articulate implications for theory, industry workforce
transformation, university curricula, and organizational leadership. We argue that,
as code becomes abundant, the durable value of the software engineer increasingly
resides in intent specification, critical judgment, and accountable oversight rather
than in the sheer volume of code produced.
\end{abstract}


\section{Introduction}
Software engineering has repeatedly been reshaped by tools that raise the level of abstraction at which practitioners work, from assemblers and high-level languages to integrated development environments, version control, and continuous integration. The current wave, driven by large language models (LLMs) and the agentic systems built upon them, differs in kind rather than degree: for the first time, tools do not merely assist human authorship but can also \emph{author}, \emph{review}, and \emph{evolve} software artifacts with limited human intervention~\cite{li2025aiteammates,roychoudhury2025trust}. The transformer architecture~\cite{vaswani2017attention} and code-trained models~\cite{chen2021codex} established the technical substrate; benchmarks such as SWE-bench~\cite{jimenez2024swebench} reframed evaluation around the resolution of real-world issues; and within a remarkably short period, autonomous coding agents moved from speculative vision to observable practice at scale~\cite{li2025aiteammates,acharya2025genai}.

Recent empirical studies further show that this transformation is not merely technical, but deeply behavioral and organizational. Hamza et al.~\cite{hamza2024human} examine human-AI collaboration in software engineering through a thematic analysis of a hands-on workshop in which 22 professional software engineers worked with ChatGPT for three hours. Their findings suggest that AI is evolving from a simple tool into a collaborative partner, improving code generation and optimization while still requiring human oversight for complex problem-solving, security, and judgment-intensive tasks. In a complementary contribution, Treude et al.~\cite{treude2025developers} propose a taxonomy of eleven interaction types between developers and AI tools, showing that AI-supported development now spans the entire lifecycle, from code completion and command-driven actions to conversational assistance. Together, these studies indicate that software engineering is no longer defined only by programming with tools, but increasingly by managing interactions with intelligent systems.

This shift forces a re-examination of two long-standing assumptions \cite{alenezi2026rethinking}. The first is that the central act of software engineering is writing code. The second is that individual coding productivity, historically and problematically proxied by output volume, is the appropriate unit of engineering value. Both assumptions are destabilized when a single developer can orchestrate multiple agents that independently generate tests, respond to review feedback, and execute multi-step workflows around the clock~\cite{li2025aiteammates}. As code generation becomes inexpensive and abundant, the binding constraint on software quality migrates from \emph{production} to \emph{intent articulation}, \emph{architectural control}, and \emph{systematic verification}~\cite{kohl2026abundant,roychoudhury2025trust}. The engineer's contribution is increasingly defined not by what they type, but by what they specify, evaluate, integrate, and remain accountable for.

Within this context, software engineering is changing from a tool-and-operator model to a manager-and-team dynamic, driven by Generative AI and Agentic AI. Generative AI supports tasks such as code completion, test generation, refactoring, and documentation, while Agentic AI can reason, plan, and carry out multi-step workflows such as fixing bugs, running tests, and opening pull requests. As a result, engineers are moving from manually producing code to orchestrating systems and guiding AI toward the right business outcomes. This evolution is also reflected in the emerging interaction patterns between developers and AI tools, which now include not only suggestions and completions, but also conversational, command-based, and workflow-oriented collaboration~\cite{treude2025developers}.

To stay effective in this environment, software engineers need stronger skills in problem decomposition, strategic planning, multi-agent orchestration, and workflow design. They also need advanced instruction design, where prompts function more like executable specifications with clear constraints and guardrails. These capabilities help ensure AI systems do not merely produce output quickly, but produce the right output in a controlled and reliable way. At the same time, engineers must become stronger reviewers and validators of AI-generated work. This includes rigorous code review, security auditing, and the ability to evaluate logic, edge cases, and system integration. A solid understanding of AI fundamentals, such as context windows, retrieval-augmented generation (RAG), and model trade-offs, is also essential so engineers can build resilient systems and treat AI as a highly capable assistant rather than a black box.

Yet the transformation is neither uniform nor frictionless. Empirical evidence indicates that although agents can dramatically accelerate code submission, their contributions are accepted less frequently than those of human developers and tend to alter fewer structural aspects of a codebase, revealing a gap between benchmark performance and real-world trust~\cite{li2025aiteammates}. In safety-critical and embedded domains, determinism, reliability, and traceability requirements make full autonomy untenable and elevate the importance of governance and human oversight~\cite{sun2026embedded}. Across the profession, scholars also warn that overreliance on AI may erode foundational skills, such as debugging, design reasoning, and critical review, which remain essential when automation fails~\cite{abrahao2025humans,mastropaolo2024riseandfall}.

These tensions motivate our central research question:

\begin{quote}
\emph{How is the integration of Generative AI and Agentic AI transforming the nature of software engineering work, and what new competencies are required for software engineers to remain effective in AI-enabled development organizations?}
\end{quote}

We address this question through a structured interpretive synthesis of a curated evidence base spanning recent roadmap articles, large-scale empirical studies of agentic contributions, domain-specific qualitative investigations, and foundational software engineering scholarship. Our contributions are fourfold:

\begin{enumerate}
\item A \textbf{paradigm characterization} that distinguishes Traditional, Generative AI-Enabled, and Agentic AI-Enabled software engineering across ten dimensions, and that locates these paradigms within the SE~1.0$\rightarrow$3.0 evolution (Section~\ref{sec:findings}).
\item An \textbf{activity transformation analysis} that classifies traditional and emerging software engineering activities as automated, augmented, or newly emerging, and projects role trajectories over the next decade (Sections~\ref{sec:findings} and \ref{sec:discussion}).
\item An original, theory-driven \textbf{Future Software Engineering Competency Framework} (FSE-CF) organizing required capabilities into five interacting categories, operationalized through a \textbf{competency matrix} (Section~\ref{sec:competency}) and a \textbf{transformation framework} (Section~\ref{sec:model}).
\item A set of \textbf{nine testable propositions} together with implications for theory, industry, education, and organizational leadership (Sections~\ref{sec:theory}--\ref{sec:future}).
\end{enumerate}

The remainder of the paper is organized as follows. Section~\ref{sec:background} reviews the background and related literature. Section~\ref{sec:method} describes the synthesis methodology. Section~\ref{sec:findings} presents the findings. Section~\ref{sec:discussion} discusses automation, augmentation, emergence, and role evolution. Sections~\ref{sec:theory} and \ref{sec:practice} develop theoretical and practical implications. Sections~\ref{sec:competency} and \ref{sec:model} present the competency framework and transformation model. Section~\ref{sec:threats} addresses threats to validity, Section~\ref{sec:future} outlines future research, and Section~\ref{sec:conclusion} concludes.

\section{Background and Literature Review}
\label{sec:background}

The proposition that artificial intelligence is reshaping software engineering can
only be assessed against a prior, and far older, understanding of what software
engineering actually is. That understanding has long resisted any reduction of the
discipline to the act of writing code. In Sommerville's classical formulation,
software engineering concerns \emph{all} aspects of software
production---requirements, design, validation, evolution, and the human processes
that bind them---rather than implementation
alone~\cite{sommerville2015software}. Brooks anticipated the consequences of this
breadth when he argued that no single technological advance would deliver
order-of-magnitude productivity gains, because much of software's difficulty is
\emph{essential}---rooted in the conceptual structure of the problem---rather than
\emph{accidental}, and therefore not amenable to better tooling
alone~\cite{brooks1975mythical}. Naur deepened the point by reframing programming as
\emph{theory building}: the durable asset of a software project is not its source
text but the shared mental model of the problem and solution domains held by the
people who build it, a model that decays when those people
leave~\cite{naur1985theory}. These foundational arguments converge on a claim that is
directly consequential for the present study: automating the production of code does
not, in itself, automate software engineering, because the discipline's hardest
problems are problems of understanding, coordination, and judgment. The same
recognition explains why measurement of developer work has steadily moved away from
single-metric proxies such as lines of code, which proved misleading, toward
multidimensional conceptions---most prominently the SPACE framework's attention to
satisfaction, performance, activity, communication, and
efficiency~\cite{forsgren2021space,abrahao2025humans}. Any serious account of AI's
effect on engineering work must therefore be evaluated not against coding throughput
but against this richer, socio-cognitive picture of the discipline.

Against that backdrop, generative AI has permeated the software development lifecycle
with unusual speed. Practitioner surveys and field reports document widespread
adoption of AI coding assistants within only a few years of their introduction,
accompanied by emergent development styles such as chat-oriented programming, ``vibe
coding,'' and agentic programming~\cite{acharya2025genai}. The reach of these
techniques now extends across requirements elicitation, design, implementation,
testing, code review, and maintenance: test cases can be synthesized from
natural-language requirements, program-repair pipelines couple generative models with
static analysis, and large language models increasingly assist vulnerability
detection and bug localization~\cite{ahmed2025journey,abrahao2025humans}. Yet the
empirical literature complicates any simple narrative of acceleration. Controlled
experimentation by Peng et al.\ found that developers granted an AI pair programmer
completed a standardized task roughly 56\% faster than a control group, with the
largest gains accruing to less-experienced
participants~\cite{peng2023copilot}. Human-centered studies, however, qualify what
such speedups mean in practice. Vaithilingam et al.\ observed that although
programmers preferred AI assistance and valued it as a starting point, it did not
reliably improve task-completion time or success rate, and frequently introduced new
difficulties in understanding, editing, and debugging generated
code~\cite{vaithilingam2022expectation}. Barke et al.\ further showed that interaction
is \emph{bimodal}---an \emph{acceleration} mode in which the developer already knows
the goal, and an \emph{exploration} mode in which the tool is used to discover
options---implying that productivity effects are contingent on task and intent rather
than uniform~\cite{barke2023grounded}. Liang et al.'s large-scale survey of 410
developers reinforces this contingency, reporting that practitioners often struggle
to obtain outputs aligned with their requirements and expectations even as they value
the reduction in keystrokes and recall
effort~\cite{liang2024survey}. The aggregate picture, as Abrah\~ao et al.\
synthesize it, is one of simultaneous tendencies: AI \emph{enhances} automation,
productivity, and quality; \emph{obsolesces} manual testing, routine coding, and
search; \emph{retrieves} natural-language specification and end-user programming; and,
when pushed too far, \emph{reverses} into pathologies of over-reliance, lost
provenance, code and feature bloat, security exposure, technical debt, and the
devaluation of craft~\cite{abrahao2025humans}.

A qualitative shift occurs as assistance gives way to autonomy. Where generative
assistants operate under continuous human control, agentic systems pursue goals with
markedly greater independence---reading codebases, planning changes, invoking tools,
running tests, and submitting pull requests~\cite{li2025aiteammates}. Roychoudhury et
al.\ argue that fielding such ``AI software engineers'' demands trust at least equal
to that earned by human-driven practice, so that the organizing concern of
programming shifts \emph{from scale to trust}~\cite{roychoudhury2025trust}. The most
direct empirical grounding for this turn is the AIDev study, which captured
456{,}535 agent-authored pull requests across 61{,}453 repositories and 47{,}303
developers from five leading agents---OpenAI Codex, GitHub Copilot, Devin, Cursor,
and Claude Code---and whose findings are at once striking and
sobering~\cite{li2025aiteammates}. Kohl and Carro generalize the implication: when
code becomes abundant, the discipline must re-center on intent articulation,
architectural control, and systematic verification, moving ``from a
production-oriented field to one centered on human judgment under
automation''~\cite{kohl2026abundant}. This re-centering is the conceptual hinge of the
present study.

A parallel literature asks what these shifts mean for the engineer's role, formation,
and accountability. Betz and Penzenstadler propose a research agenda for a ``new
generation'' of software engineers and a competency--method--technology scaffold
spanning long-term thinking, systems and complexity thinking, sustainability
assessment, and ethics as a practical skill~\cite{betz2025responsibility}. Abrah\~ao
et al.\ foreground human factors---trust, autonomy, creativity, and the dynamics of
hybrid human--AI teams---and anticipate the growth of ``citizen'' developers
alongside professionals~\cite{abrahao2025humans}. Marlowe et al.\ contend that
software engineering education must intertwine process thinking, systems theory, and
ethical engagement as technical responsibilities shift
``upward''~\cite{marlowe2026curriculum}, while Goth et al.\ demonstrate the practical
value of an explicit, progression-based competency framework for hiring, training,
and curriculum design among research software
engineers~\cite{goth2025rse}. From a management vantage, Bruhin et al.\ analyze
generative AI's impact across the People--Process--Technology pillars and propose
strategic organizational actions~\cite{bruhin2024genai}, and Alenezi and Akour review
AI-driven innovations and future directions across the
lifecycle~\cite{alenezi2025innovations}. Cutting across these strands is a distinct
engineering challenge that the rise of AI collaborators makes mainstream: building
systems that \emph{contain} machine-learning components was already known to incur
``hidden technical debt'' and to demand new practices for data, models, and
monitoring~\cite{sculley2015debt,amershi2019se4ml}; as AI migrates from a feature
inside systems to a collaborator that produces systems, these data-centric and
observability concerns move from a specialized sub-discipline into baseline
competence.

Taken together, the literature exhibits both convergence and a conspicuous gap. It
converges on the message that the engineer of the AI era is less a lone builder than a
disciplined collaborator with machines---a curator of intelligent systems who manages
intent, context, orchestration, verification, and governance while retaining enough
foundational depth to intervene when tools
fail~\cite{abrahao2025humans,roychoudhury2025trust,betz2025responsibility}. Yet the
evidence remains fragmented across methodological silos: controlled productivity
experiments~\cite{peng2023copilot}, human-centered usability
studies~\cite{vaithilingam2022expectation,barke2023grounded,liang2024survey},
large-scale repository
analyses~\cite{li2025aiteammates}, and conceptual or educational
agendas~\cite{betz2025responsibility,marlowe2026curriculum} seldom speak to one
another, and few integrate the socio-cognitive foundations of the discipline with the
emerging empirics of agentic practice. The transformation is consequently described
in fragments rather than as a coherent reconfiguration of work. This study addresses
that gap by synthesizing these strands into an integrated account of human--AI
collaboration in software engineering, asking how the locus of engineering
competence and value is being redistributed as autonomous agents assume an
increasing share of production---and what foundational depth, judgment, and
governance human engineers must retain for that collaboration to remain trustworthy.

\section{Research Methodology}
\label{sec:method}

\subsection{Research Design}
This study is a \emph{theory-building structured interpretive synthesis}. Because the
phenomenon---the integration of GenAI and Agentic AI into engineering work---is
recent, fast-moving, and documented across heterogeneous sources (peer-reviewed
journal roadmaps, archival empirical studies, and domain investigations), a
configurative synthesis that interprets and integrates evidence is more appropriate
than an aggregative meta-analysis of homogeneous trials. Our design follows
established guidance for qualitative and multivocal evidence synthesis in software
engineering: an explicit corpus, transparent inclusion criteria, thematic coding, and
a traceable path from evidence to framework. The synthesis is interpretive and
generative: its output is a conceptual framework and a set of falsifiable
propositions, not a pooled effect size.

\subsection{Evidence Corpus and Inclusion Criteria}
The corpus combines a curated set of recent core sources (2024--2026) addressing the
research question directly with a layer of foundational references that anchor
constructs (e.g., productivity, complexity, ML-system engineering). Sources were
included if they (i) addressed the impact of generative or agentic AI on software
engineering work, roles, or competencies; (ii) were either peer-reviewed or
archival manuscripts authored by identifiable researchers at recognized institutions;
and (iii) contributed empirical observations, structured roadmaps, or competency
models. We excluded vendor marketing material, opinion pieces lacking argument or
evidence, and sources whose provenance could not be verified. Every citation was
verified against its primary record (publisher DOI or archival identifier).
Table~\ref{tab:corpus} summarizes the core corpus by type and contribution.

\begin{table*}[t]
\centering
\caption{Core evidence corpus and primary contribution to the synthesis.}
\label{tab:corpus}
\footnotesize
\begin{tabularx}{\textwidth}{L{4.7cm} L{2.5cm} Y}
\toprule
\textbf{Source} & \textbf{Type} & \textbf{Primary contribution to this synthesis}\\
\midrule
Li, Zhang \& Hassan~\cite{li2025aiteammates} & Large-scale empirical (archival) &
Empirical evidence of autonomous coding agents in the wild (AIDev: 456{,}535 PRs);
SE~1.0$\rightarrow$3.0 paradigm articulation; acceptance, speed, and complexity gaps.\\
Abrah\~ao et al.~\cite{abrahao2025humans} & Journal roadmap (TOSEM) &
Human-centric SE roadmap; disruptive-effects tetrad; productivity, trust, autonomy,
creativity; evolution of the development context toward hybrid teams.\\
Ahmed et al.~\cite{ahmed2025journey} & Journal roadmap (TOSEM) &
AI-for-SE technical and organizational challenges across the lifecycle; 2030 research
horizon; human--AI collaboration in teams.\\
Betz \& Penzenstadler~\cite{betz2025responsibility} & Journal roadmap (TOSEM) &
Role, responsibility, ethics, and sustainability; competency--method--technology
scaffold; ten research challenges.\\
Roychoudhury et al.~\cite{roychoudhury2025trust} & Position/opinion (archival) &
``Programming with trust''; integration of LLM agents with analysis tools; shift of
focus from scale to verifiable trust.\\
Acharya~\cite{acharya2025genai} & Survey/synthesis (archival) &
New paradigms (CHOP, vibe coding, agentic programming); orchestration, MCP,
multi-agent systems; emerging roles and skills.\\
Sun \& Staron~\cite{sun2026embedded} & Qualitative (archival) &
Agentic pipelines in embedded/safety-critical SE; eleven emerging practices and
fourteen challenges; governance, determinism, traceability.\\
Kohl \& Carro~\cite{kohl2026abundant} & Position (ICSE-FoSE) &
Code-abundance thesis; redefinition of SE around intent, architectural control, and
verification.\\
Marlowe et al.~\cite{marlowe2026curriculum} & Position (JIDPS) &
SE education redesign: process thinking, systems theory, ethics, soft skills,
lifelong learning.\\
Goth et al.~\cite{goth2025rse} & Community/empirical (F1000) &
Foundational, progression-based competency framework for research software engineers.\\
Bruhin et al.~\cite{bruhin2024genai} & Empirical/strategy (ICIS) &
People--Process--Technology analysis; strategic managerial actions for GenAI adoption.\\
Mastropaolo et al.~\cite{mastropaolo2024riseandfall} & Position (archival) &
Integration of AI into SE while preserving human creativity and craft.\\
\bottomrule
\end{tabularx}
\end{table*}

\subsection{Analysis Procedure}
Analysis proceeded in four passes. First, in \emph{open coding}, we extracted
claims, observations, and recommendations relevant to work transformation and
competencies. Second, in \emph{axial coding}, we grouped codes into candidate themes
(e.g., \emph{orchestration}, \emph{verification}, \emph{governance}, \emph{intent
specification}, \emph{deskilling risk}). Third, in \emph{framework synthesis}, we
mapped themes onto two organizing dimensions---(a)~the paradigm continuum
(Traditional$\rightarrow$GenAI$\rightarrow$Agentic) and (b)~a competency taxonomy
refined iteratively against the corpus---and resolved overlaps by constant
comparison. Fourth, in \emph{theoretical integration}, we derived propositions
intended to be falsifiable by future empirical work. To strengthen credibility, each
framework element is traceable to at least one corpus source, and triangulation was
sought across source types (empirical, roadmap, and domain study) before a claim was
treated as well supported.

\subsection{Validity Safeguards}
We treated archival manuscripts as evidence of \emph{direction and mechanism} rather
than settled effect magnitudes, and we cross-checked their claims against
peer-reviewed roadmaps and foundational theory. Where the corpus disagreed (for
example, on the degree of autonomy advisable in safety-critical contexts), we
preserved the disagreement rather than averaging it away. Threats arising from this
design are analyzed explicitly in Section~\ref{sec:threats}.

\section{Findings and Analysis}
\label{sec:findings}

\subsection{Three Coexisting Paradigms}
Our first finding is that contemporary practice is best understood not as a single,
uniform state but as three \emph{coexisting} paradigms whose relative weight is
gradually shifting rather than abruptly replacing one another. Crucially, these
paradigms are not mutually exclusive: individual teams, projects, and even single
workflows frequently blend elements of each, and the boundaries between them are
porous rather than sharply delineated. Building on the
SE~1.0$\rightarrow$3.0 articulation~\cite{li2025aiteammates}, we consolidate this
transition into three working paradigms and compare them systematically across ten
dimensions in Table~\ref{tab:paradigms}, allowing the distinctive characteristics and
trade-offs of each to be examined side by side. To complement this
cross-sectional comparison and convey the directional trajectory of the shift, the
progression is further summarized as a maturity ladder in
Table~\ref{tab:maturity}, which positions the paradigms as successive stages of
increasing autonomy and organizational integration. Taken together, these two
representations characterize both the structure of the present landscape and the
direction in which practice appears to be evolving.

\begin{table}[t]
\centering
\caption{Paradigm maturity ladder (consolidating SE~1.0--3.0).}
\label{tab:maturity}
\footnotesize
\begin{tabularx}{\columnwidth}{L{1.4cm} L{3cm} Y}
\toprule
\textbf{Stage} & \textbf{Label} & \textbf{Human role}\\
\midrule
SE~1.0 & Classical & Manual authorship; explicit rules; deterministic flows.\\
SE~1.5 & Predictive coding & Accept/ignore token-level suggestions in real time.\\
SE~2.0 & AI-assisted (GenAI) & Provide intent; review and integrate AI output.\\
SE~3.0 & Agentic & Orchestrate goal-driven agents; set permissions; verify and govern.\\
\bottomrule
\end{tabularx}
\end{table}

\begin{table*}[t]
\centering
\caption{Comparison of Traditional, Generative AI-Enabled, and Agentic AI-Enabled software engineering across ten dimensions.}
\label{tab:paradigms}
\footnotesize
\begin{tabularx}{\textwidth}{L{3.0cm} Y Y Y}
\toprule
\textbf{Dimension} & \textbf{Traditional SE} & \textbf{Generative AI-Enabled SE} & \textbf{Agentic AI-Enabled SE}\\
\midrule
Locus of authorship & Human writes code & Human writes; AI co-authors fragments & Agents author, test, and submit changes\\
Primary human role & Implementer/designer & Director and reviewer of suggestions & Orchestrator, verifier, and accountable owner\\
Unit of work & Functions, modules, commits & Prompt--response increments & Goals/tasks delegated to agents\\
Dominant productivity proxy & Output volume, velocity & Accelerated velocity (SPACE-aware) & Throughput gated by verification capacity\\
Verification \& validation & Human testing/review & Human review of generated code & Multi-layer V\&V of high-throughput agent output\\
Team structure & Human teams & Humans + assistants & Hybrid human--agent teams (incl.\ review bots)\\
Knowledge artifact & Code + design docs & Code + prompts/context & Intent specs, policies, traces, and provenance\\
Principal risk & Defects, schedule, scope & Plausible-but-wrong code; bloat; bias & Trust gaps; opaque autonomy; accountability voids\\
Governance posture & Process and standards & Usage guidelines; IP/security review & Explicit autonomy policies; auditability; oversight\\
Critical competency & Coding and design & Prompt/context engineering; critical review & Orchestration; V\&V; governance; systems thinking\\
\bottomrule
\end{tabularx}
\end{table*}

\subsection{Empirical Grounding: Speed Without Settled Trust}
Our second finding, grounded chiefly in AIDev, is that agentic contribution is
\emph{already real and large-scale}, yet \emph{trust lags
capability}~\cite{li2025aiteammates}. Three observations are pivotal. First, agents
can compress timelines dramatically: a single developer was observed producing 164
agent-authored pull requests in three days---comparable to 176 human-authored pull
requests over the prior three years---and certain agents close pull requests up to an
order of magnitude faster than humans~\cite{li2025aiteammates}. Second, despite this
speed, agent-authored pull requests are \emph{less likely to be merged} than
human-authored ones, particularly for complex tasks such as feature development and
bug fixing, and they tend to alter fewer structural aspects of code (e.g., lower rates
of cyclomatic-complexity change)~\cite{li2025aiteammates}. Third, review is itself
being reconfigured: provider-aligned ``closed loops'' emerge in which an agent's
output is reviewed predominantly by a review bot from the same provider, streamlining
workflows but risking reinforcement of provider-specific blind spots; notably,
trusted security review remains substantially rule-based rather than
AI-driven~\cite{li2025aiteammates}. The synthesis of these observations is that the
binding constraint is shifting from \emph{how fast code can be produced} to
\emph{how reliably it can be trusted, integrated, and maintained}---exactly the
argument advanced from first principles by Roychoudhury et
al.~\cite{roychoudhury2025trust} and Kohl and Carro~\cite{kohl2026abundant}.

\subsection{The Migration of Effort}
Third, effort is migrating along the value chain. As generation is automated, the
relative weight of \emph{specification} (translating ambiguous business needs into
precise, machine-actionable intent), \emph{verification} (challenging plausible
output), \emph{integration} (fitting contributions into a coherent architecture), and
\emph{governance} (ensuring alignment with quality, security, and policy) rises. This
is consistent across the corpus: Acharya documents new orchestration and
prompt-orchestration practices~\cite{acharya2025genai}; Sun and Staron find embedded
teams reorganizing workflows, roles, and toolchains around agentic pipelines while
insisting that agents remain collaborators rather than autonomous actors in
safety-critical settings~\cite{sun2026embedded}; and Betz and Penzenstadler locate the
engineer's expanded responsibility in long-term, systemic, and ethical
reasoning~\cite{betz2025responsibility}.

\subsection{Activity Transformation: Automated, Augmented, Emerging}
Fourth, the impact on specific activities is uneven. We classify representative
software engineering activities into those primarily being \emph{automated}, those
being \emph{augmented}, and those newly \emph{emerging}, with supporting evidence, in
Table~\ref{tab:activities}. The pattern is instructive: routine, well-specified, and
boilerplate-heavy tasks are most automatable; judgment-intensive, context-dependent,
and accountability-bearing tasks are augmented rather than replaced; and an entirely
new layer of work---orchestration, context engineering, verification of AI output,
and AI governance---emerges to manage the automation itself.

\begin{table*}[t]
\centering
\caption{Transformation of software engineering activities: automated, augmented, and emerging.}
\label{tab:activities}
\footnotesize
\begin{tabularx}{\textwidth}{L{2.4cm} Y L{4.6cm}}
\toprule
\textbf{Mode} & \textbf{Representative activities} & \textbf{Evidence / rationale}\\
\midrule
\textbf{Automated} \newline (largely machine-performed) &
Boilerplate and scaffolding; routine unit-test generation; documentation drafts;
syntactic refactoring; first-pass code completion; routine bug localization and patch
suggestion. &
Generation from prompts and specifications; program repair with static analysis; bug
localization~\cite{abrahao2025humans,acharya2025genai,li2025aiteammates}.\\
\addlinespace
\textbf{Augmented} \newline (human judgment amplified) &
Requirements elicitation and disambiguation; architectural design; complex debugging;
code review; security analysis; estimation; stakeholder communication. &
AI accelerates familiar work and probes design spaces, but humans retain control,
trust calibration, and integration of mental
models~\cite{abrahao2025humans,ahmed2025journey,roychoudhury2025trust}.\\
\addlinespace
\textbf{Emerging} \newline (new work created by AI) &
Intent/context engineering; agent orchestration and workflow design; multi-layer V\&V
of agent output; AI governance and autonomy policy; provenance/audit; model monitoring
and observability; AI-native architecture. &
New roles and skills, orchestration and MCP integration, review-triage and governance
of high-throughput agentic
contributions~\cite{acharya2025genai,kohl2026abundant,sun2026embedded,li2025aiteammates}.\\
\bottomrule
\end{tabularx}
\end{table*}

\subsection{A Persistent Need for Foundational Depth}
Fifth, and counterbalancing the automation narrative, the corpus is consistent that
foundational engineering depth remains necessary---arguably more so. A ``developer in
the loop'' is needed when automation breaks down, and debugging or repairing
AI-generated code at scale introduces new cognitive
demands~\cite{abrahao2025humans}. Mastropaolo et al.\ emphasize preserving human
creativity and craft amid integration~\cite{mastropaolo2024riseandfall}. The clear
implication is a \emph{barbell} skill profile: deep foundations (algorithms, design,
debugging, security) combined with new AI-collaboration capabilities, with the
middle---routine production---increasingly delegated.

\section{Discussion: Automation, Augmentation, Emergence, and Role Evolution}
\label{sec:discussion}

\subsection{What Is Being Automated---and Why It Is Not Enough}
The activities most amenable to automation share three properties: they are
well-specified, locally scoped, and weakly coupled to organizational accountability.
Automating them yields genuine velocity gains but does not, by itself, deliver
trustworthy systems, because the residual work---deciding \emph{what} to build,
\emph{whether} the output is correct, and \emph{who} is answerable for it---is
precisely the work that resists automation~\cite{roychoudhury2025trust,kohl2026abundant}.
This explains the empirical paradox of fast-but-less-merged agentic pull
requests~\cite{li2025aiteammates}: speed in production does not compensate for
deficits in trust and integration.

\subsection{What Is Being Augmented---and the Trust-Calibration Problem}
Augmentation is double-edged. AI can be used in an \emph{acceleration} mode (speeding
familiar work) or an \emph{exploration} mode (probing unfamiliar design spaces), and
realized productivity depends on balancing rapid generation against validation and
trust overhead~\cite{li2025aiteammates,abrahao2025humans}. The central human skill in
augmentation is therefore \emph{trust calibration}: neither blind acceptance nor
reflexive rejection, but evidence-weighted judgment about when to rely on, verify, or
override AI output. Mis-calibration in either direction is costly---over-trust
propagates plausible-but-wrong code, while under-trust forfeits the gains.

\subsection{What Is Emerging---a New Layer of Engineering Work}
The emergent layer is the most consequential for competency demand. Orchestration
turns one engineer into the supervisor of many concurrent agentic workstreams,
changing the cadence and locus of decision-making and inflating review
load~\cite{li2025aiteammates,acharya2025genai}. Context engineering---curating the
information, constraints, and tools an agent can access---becomes a determinant of
output quality. Verification scales from individual review to the design of
\emph{verification systems} (oracles, test strategies, and policy checks) capable of
handling high-throughput contributions~\cite{roychoudhury2025trust}. And governance
moves from a compliance afterthought to a first-class engineering activity:
specifying when agents may act autonomously, how contributions are validated, and who
holds accountability for defects~\cite{li2025aiteammates,sun2026embedded}.

\subsection{Role Evolution Over the Next Decade}
Synthesizing these dynamics, we project the evolution of several roles
(Table~\ref{tab:roles}). The throughline is a shift in the center of gravity of
engineering value from \emph{authorship} toward \emph{orchestration, verification, and
accountable judgment}---accompanied by the rise of genuinely new roles (e.g., AI
workflow/orchestration engineer, AI governance/assurance lead) and the elevation of
architecture and security from specialist concerns to broadly held
competencies~\cite{abrahao2025humans,bruhin2024genai,sun2026embedded}. We caution
that these are \emph{plausible trajectories} conditioned on current evidence, not
deterministic forecasts.

\begin{table*}[t]
\centering
\caption{Plausible role trajectories over the next decade.}
\label{tab:roles}
\footnotesize
\begin{tabularx}{\textwidth}{L{3.6cm} Y Y}
\toprule
\textbf{Role} & \textbf{Traditional emphasis} & \textbf{Projected AI-enabled emphasis}\\
\midrule
Software developer & Implementing features in code & Specifying intent, orchestrating agents, verifying and integrating output\\
Senior/staff engineer & Technical leadership; complex implementation & System-level judgment; verification strategy; mentoring trust calibration\\
Software architect & Structural design of components & Architecture of AI-native, human-in-the-loop socio-technical systems\\
QA / test engineer & Writing and executing tests & Designing verification systems and oracles for high-throughput agent output\\
Security engineer & Specialist review and hardening & Embedded, lifecycle-wide security and privacy oversight of AI artifacts\\
\emph{(New)} AI workflow engineer & --- & Designing, orchestrating, and observing multi-agent pipelines\\
\emph{(New)} AI governance lead & --- & Autonomy policy, accountability, provenance, and responsible-AI assurance\\
\bottomrule
\end{tabularx}
\end{table*}

\section{Theoretical Implications}
\label{sec:theory}
Our synthesis carries several implications for theory. First, it motivates an
\textbf{intent-centric reconceptualization of software engineering value}. If Naur's
``theory building''~\cite{naur1985theory} located the durable asset in developers'
shared understanding, the AI era sharpens this claim: when code is cheap, the theory
of the problem---expressed as intent, constraints, and acceptance criteria---becomes
the scarce, value-bearing artifact, and Brooks's distinction between essential and
accidental complexity~\cite{brooks1975mythical} maps cleanly onto the boundary between
what resists and what yields to automation.

Second, it calls for \textbf{extending socio-technical systems theory to hybrid
human--agent teams}. Classical accounts assume human actors; agentic teammates that
make micro-decisions, review one another, and operate continuously violate that
assumption~\cite{li2025aiteammates}. Constructs such as trust, coordination, and
team velocity require redefinition when some teammates are non-human, and provider
ecosystems can introduce systematic correlation (and bias) across both authorship and
review~\cite{li2025aiteammates}.

Third, it implies that \textbf{productivity theory must absorb verification and trust
as first-class dimensions}. The SPACE framework~\cite{forsgren2021space} already
resists single-metric reduction; our findings suggest an explicit verification-cost
and trust-calibration dimension, since net throughput is gated not by generation speed
but by the capacity to validate output~\cite{li2025aiteammates,roychoudhury2025trust}.

Fourth, it positions \textbf{governance as constitutive of engineering rather than
external to it}. When autonomy is delegated, accountability does not
disappear---it concentrates~\cite{betz2025responsibility,sun2026embedded}. A theory of
AI-enabled software engineering must therefore treat governance competencies as part
of the engineering core, not as a managerial overlay.

\section{Practical and Industry Implications}
\label{sec:practice}

\subsection{For Organizational Leaders}
Leaders should treat AI adoption as a \emph{socio-technical transformation} across
People, Process, and Technology, not as tool procurement~\cite{bruhin2024genai}.
Concretely: (i)~invest in \emph{verification capacity}---review tooling, test
infrastructure, and reviewer time---so that generation gains are not lost to
integration debt~\cite{li2025aiteammates}; (ii)~establish \emph{autonomy and
accountability policies} specifying when agents may act, how their output is
validated, and who is answerable for defects~\cite{li2025aiteammates,sun2026embedded};
(iii)~avoid \emph{provider monoculture} in agent and reviewer selection to mitigate
correlated blind spots~\cite{li2025aiteammates}; and (iv)~measure productivity with
multidimensional, verification-aware metrics rather than output
volume~\cite{forsgren2021space}.

\subsection{For Workforce Transformation}
Reskilling should target the \emph{barbell} profile: deepen foundations while building
AI-collaboration competencies. Organizations should create explicit progression paths
for emerging roles (orchestration, governance, AI-native architecture), recognize and
reward trust calibration and verification work, and guard against deskilling by
preserving opportunities for engineers to exercise core judgment rather than only
supervising automation~\cite{abrahao2025humans,mastropaolo2024riseandfall}.

\subsection{For Universities and Curriculum Design}
Curricula should intertwine process thinking, systems theory, ethics, and soft skills
with foundational rigor, and treat prompt/context engineering, verification of AI
output, and AI governance as examinable
competencies~\cite{marlowe2026curriculum,betz2025responsibility}. A
\emph{progression-based competency model}, as demonstrated for research software
engineers~\cite{goth2025rse}, offers a practical template: define foundational
competencies, articulate proficiency levels, and map courses and assessments onto
them. Crucially, students must still learn to build and debug systems unaided, so that
AI fluency rests on, rather than substitutes for, engineering judgment.

\section{The Future Software Engineering Competency Framework (FSE-CF)}
\label{sec:competency}
\begin{figure*}[t]
\centering
\resizebox{\textwidth}{!}{%
\begin{tikzpicture}[node distance=3mm]
\node[cat, fill=black!12, minimum width=3.0cm] (core) {Future SE\\Competency (FSE-CF)};

\node[cat, above left=10mm and 18mm of core] (tech) {Technical};
\node[cat, above right=10mm and 18mm of core] (cog) {Cognitive};
\node[cat, right=26mm of core] (soc) {Socio-technical};
\node[cat, below right=10mm and 18mm of core] (gov) {Governance};
\node[cat, below left=10mm and 18mm of core] (org) {Organizational};

\draw[flow] (core) -- (tech);
\draw[flow] (core) -- (cog);
\draw[flow] (core) -- (soc);
\draw[flow] (core) -- (gov);
\draw[flow] (core) -- (org);

\node[leafn, left=4mm of tech, text width=3.5cm] (techL)
{Prompt engineering; context engineering; AI workflow design; AI-native architecture;
data-centric engineering; security \& privacy oversight; model monitoring \&
observability};
\draw[dflow] (tech) -- (techL);

\node[leafn, above=4mm of cog, text width=3.7cm] (cogL)
{Critical evaluation \& judgment; systems thinking; verification \& validation of AI
output; intent specification; abstraction management; metacognition \& lifelong
learning};
\draw[dflow] (cog) -- (cogL);

\node[leafn, right=4mm of soc, text width=3.3cm] (socL)
{Human--AI collaboration; agent orchestration; stakeholder translation; coordination
in hybrid human--agent teams; trust calibration};
\draw[dflow] (soc) -- (socL);

\node[leafn, below=4mm of gov, text width=3.6cm] (govL)
{AI governance; responsible AI \& ethics; accountability \& provenance; risk
management; standards \& compliance};
\draw[dflow] (gov) -- (govL);

\node[leafn, left=4mm of org, text width=3.4cm] (orgL)
{Workflow \& process (re)design; change management; capability building; sustainability;
portfolio \& role design};
\draw[dflow] (org) -- (orgL);
\end{tikzpicture}%
}
\caption{The Future Software Engineering Competency Framework (FSE-CF): five
interacting categories and their constituent competencies, synthesized from the
corpus.}
\label{fig:taxonomy}
\end{figure*}
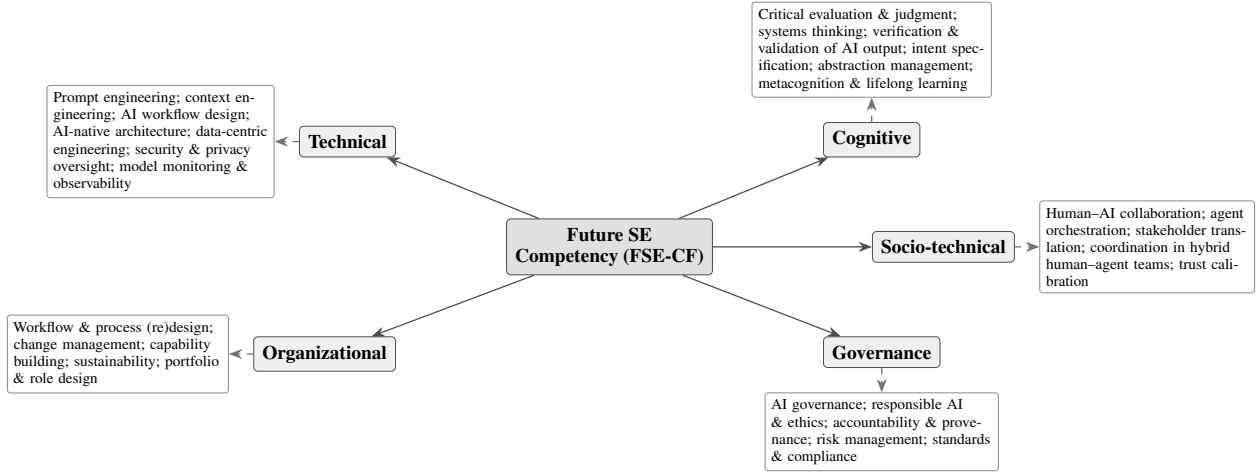

We now present the paper's principal constructive contribution: an original
competency framework synthesized from the corpus. The FSE-CF organizes the
capabilities required of effective engineers in AI-enabled organizations into five
interacting categories---\emph{technical}, \emph{cognitive}, \emph{socio-technical},
\emph{governance}, and \emph{organizational}. The categories are interacting rather
than independent: technical capability is exercised through cognitive judgment,
within socio-technical collaboration, under governance constraints, and inside
organizational structures. Figure~\ref{fig:taxonomy} depicts the taxonomy;
Table~\ref{tab:framework} defines each competency and links it to supporting evidence;
and Table~\ref{tab:matrix} provides an operational \emph{competency matrix}
classifying each competency by transformation type, AI-substitutability, strategic
criticality, and observable proficiency indicators.

\begin{table*}[t]
\centering
\caption{The FSE-CF: competency definitions and supporting evidence.}
\label{tab:framework}
\footnotesize
\begin{tabularx}{\textwidth}{L{2.2cm} L{3.4cm} Y c}
\toprule
\textbf{Category} & \textbf{Competency} & \textbf{Definition (what the engineer must be able to do)} & \textbf{Evidence}\\
\midrule
\multirow{4}{=}{\textbf{Technical}}
 & Prompt \& context engineering & Frame tasks, constraints, and context so that AI/agents produce reliable, on-target output. & \cite{acharya2025genai}\\
 & AI workflow design & Compose and configure multi-agent pipelines and toolchains into coherent processes. & \cite{acharya2025genai,sun2026embedded}\\
 & AI-native architecture & Design systems with AI components, control loops, data flows, and human-oversight points. & \cite{kohl2026abundant,amershi2019se4ml}\\
 & Data-centric eng.\ \& observability & Engineer data quality and monitor models/agents for drift, regressions, and risk. & \cite{sculley2015debt,amershi2019se4ml}\\
\midrule
\multirow{4}{=}{\textbf{Cognitive}}
 & Critical evaluation \& judgment & Challenge plausible-but-wrong output; weigh evidence; decide when to trust or override. & \cite{abrahao2025humans,roychoudhury2025trust}\\
 & Systems thinking & Reason about whole-system behavior, dependencies, and long-term consequences. & \cite{betz2025responsibility,marlowe2026curriculum}\\
 & V\&V of AI output & Design oracles, tests, and checks that validate high-throughput contributions. & \cite{roychoudhury2025trust,li2025aiteammates}\\
 & Intent specification & Translate ambiguous needs into precise, machine-actionable intent and acceptance criteria. & \cite{kohl2026abundant,acharya2025genai}\\
\midrule
\multirow{3}{=}{\textbf{Socio-technical}}
 & Human--AI collaboration & Form accurate mental models of AI behavior and adapt workflows around them. & \cite{abrahao2025humans,ahmed2025journey}\\
 & Agent orchestration & Coordinate concurrent agentic workstreams while preserving oversight and control. & \cite{li2025aiteammates,acharya2025genai}\\
 & Trust calibration & Maintain evidence-weighted reliance---neither over-trust nor reflexive rejection. & \cite{li2025aiteammates,roychoudhury2025trust}\\
\midrule
\multirow{3}{=}{\textbf{Governance}}
 & AI governance \& autonomy policy & Define when agents may act, how output is validated, and who is accountable. & \cite{li2025aiteammates,sun2026embedded}\\
 & Responsible AI \& ethics & Identify and mitigate bias, opacity, and harm as a practical engineering skill. & \cite{betz2025responsibility,abrahao2025humans}\\
 & Security \& privacy oversight & Embed lifecycle-wide security and privacy assurance over AI-generated artifacts. & \cite{li2025aiteammates,sun2026embedded}\\
\midrule
\multirow{2}{=}{\textbf{Organizational}}
 & Process (re)design \& change mgmt. & Redesign workflows, roles, and metrics for hybrid human--agent teams. & \cite{bruhin2024genai,sun2026embedded}\\
 & Capability building \& sustainability & Cultivate progression paths and sustainable, responsible adoption. & \cite{goth2025rse,betz2025responsibility}\\
\bottomrule
\end{tabularx}
\end{table*}

\begin{table*}[t]
\centering
\caption{Competency matrix: transformation type, AI-substitutability of the human competency, strategic criticality, and proficiency indicators.}
\label{tab:matrix}
\footnotesize
\begin{tabularx}{\textwidth}{L{3.6cm} L{1.9cm} L{2.0cm} L{1.9cm} Y}
\toprule
\textbf{Competency} & \textbf{Transformation} & \textbf{AI-substitut. (human skill)} & \textbf{Strategic criticality} & \textbf{Observable proficiency indicator}\\
\midrule
Prompt \& context engineering & New & Medium & High & Reliably elicits correct output; designs reusable context scaffolds.\\
AI workflow design & New & Low & High & Builds maintainable multi-agent pipelines with oversight points.\\
AI-native architecture & Reconfigured & Low & High & Produces architectures with explicit human-in-the-loop control.\\
Data-centric eng.\ \& observability & Elevated & Medium & High & Detects drift/regressions; instruments models and agents.\\
Critical evaluation \& judgment & Elevated & Very low & Very high & Catches plausible-but-wrong output; justifies trust decisions.\\
Systems thinking & Elevated & Very low & Very high & Anticipates whole-system and long-term effects of changes.\\
V\&V of AI output & Elevated & Low & Very high & Designs oracles/tests scaling to high-throughput contributions.\\
Intent specification & Elevated & Low & Very high & Converts ambiguity into testable acceptance criteria.\\
Human--AI collaboration & New & Very low & High & Adapts workflow to AI strengths/limits; effective hand-offs.\\
Agent orchestration & New & Low & High & Manages concurrent agents within review/oversight capacity.\\
Trust calibration & New & Very low & Very high & Avoids both over-trust and under-trust under evidence.\\
AI governance \& autonomy policy & New & Very low & Very high & Specifies autonomy limits, validation, and accountability.\\
Responsible AI \& ethics & Elevated & Very low & High & Identifies and mitigates bias/opacity during development.\\
Security \& privacy oversight & Elevated & Low & Very high & Embeds assurance across the AI-augmented lifecycle.\\
Process (re)design \& change mgmt. & New & Very low & High & Redesigns roles, workflows, and verification-aware metrics.\\
Capability building \& sustainability & Elevated & Very low & High & Builds progression paths; sustains responsible adoption.\\
\bottomrule
\end{tabularx}
\end{table*}

The matrix yields a clear pattern: the competencies with the \emph{lowest}
AI-substitutability and the \emph{highest} strategic criticality are cognitive and
governance competencies---critical judgment, systems thinking, trust calibration, and
autonomy/accountability policy. This inverts the traditional hierarchy in which coding
proficiency sat at the center; in AI-enabled organizations, coding becomes a
necessary but increasingly delegated competency, while judgment and governance become
the differentiators.

\section{A Transformation Framework and Conceptual Model}
\label{sec:model}
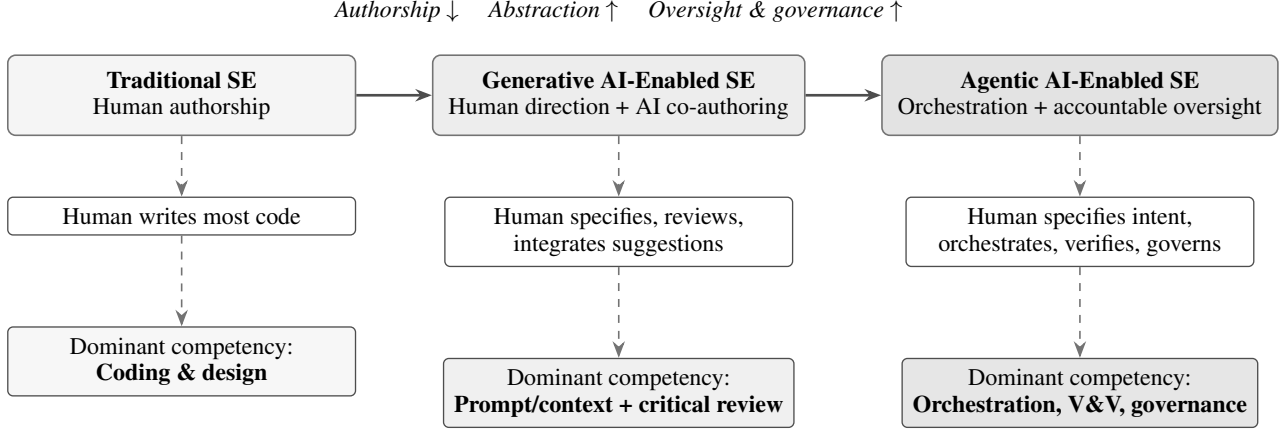
\begin{figure*}[t]
\centering
\begin{tikzpicture}[node distance=6mm]
\node[lane, fill=black!4, minimum width=4.6cm, minimum height=1.0cm] (p1) {Traditional SE\\\footnotesize\mdseries Human authorship};
\node[lane, right=10mm of p1, fill=black!7, minimum width=4.6cm, minimum height=1.0cm] (p2) {Generative AI-Enabled SE\\\footnotesize\mdseries Human direction + AI co-authoring};
\node[lane, right=10mm of p2, fill=black!11, minimum width=4.6cm, minimum height=1.0cm] (p3) {Agentic AI-Enabled SE\\\footnotesize\mdseries Orchestration + accountable oversight};

\draw[flow, line width=1pt] (p1) -- (p2);
\draw[flow, line width=1pt] (p2) -- (p3);

\node[box, below=8mm of p1, minimum width=4.6cm] (h1) {Human writes most code};
\node[box, below=8mm of p2, minimum width=4.6cm] (h2) {Human specifies, reviews,\\ integrates suggestions};
\node[box, below=8mm of p3, minimum width=4.6cm] (h3) {Human specifies intent,\\ orchestrates, verifies, governs};
\draw[dflow] (p1) -- (h1);
\draw[dflow] (p2) -- (h2);
\draw[dflow] (p3) -- (h3);

\node[box, below=12mm of h1, minimum width=4.6cm, fill=black!3] (c1) {Dominant competency:\\ \textbf{Coding \& design}};
\node[box, below=12mm of h2, minimum width=4.6cm, fill=black!6] (c2) {Dominant competency:\\ \textbf{Prompt/context + critical review}};
\node[box, below=12mm of h3, minimum width=4.6cm, fill=black!10] (c3) {Dominant competency:\\ \textbf{Orchestration, V\&V, governance}};
\draw[dflow] (h1) -- (c1);
\draw[dflow] (h2) -- (c2);
\draw[dflow] (h3) -- (c3);

\node[font=\footnotesize\itshape, above=3mm of p2] (lbl) {Authorship $\downarrow$ \quad Abstraction $\uparrow$ \quad Oversight \& governance $\uparrow$};
\end{tikzpicture}
\caption{Transformation framework. Across paradigms, the human contribution migrates
from authorship to orchestration and accountable oversight, while the dominant
competency shifts from coding toward cognitive and governance capabilities.}
\label{fig:transform}
\end{figure*}

\begin{figure}[t]
\centering
\begin{tikzpicture}[node distance=8mm and 9mm]
\node[box, fill=black!8] (intent) {Intent \&\\Context\\(human)};
\node[box, right=of intent] (gen) {Generation\\(agents)};
\node[box, below=of gen] (vv) {Verification\\\& Validation};
\node[box, left=of vv] (integ) {Integration};
\node[box, below=14mm of intent, fill=black!8] (gov) {Governance \&\\human-in-the-loop};

\draw[flow] (intent) -- (gen);
\draw[flow] (gen) -- (vv);
\draw[flow] (vv) -- (integ);
\draw[flow] (integ) -- (gov);
\draw[flow] (gov) -- (intent);
\draw[dflow] (gov.east) to[bend right=18] (vv.south);
\node[font=\scriptsize\itshape, right=1mm of vv, text width=2.0cm] {trust is produced\\by the loop};
\end{tikzpicture}
\caption{Conceptual model of AI-enabled software engineering as a closed
socio-technical loop. Human intent and context drive agentic generation; output is
verified, validated, and integrated; a governance layer with a persistent
human-in-the-loop checkpoint feeds learning back into intent.}
\label{fig:loop}
\end{figure}
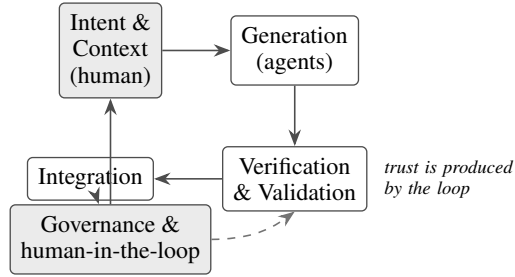

We integrate the findings into two complementary models. The \textbf{transformation
framework} (Figure~\ref{fig:transform}) shows the movement across the three paradigms
and the corresponding migration of the human contribution---from authorship, through
direction, to orchestration and accountable oversight---alongside the rising salience
of cognitive and governance competencies. The \textbf{conceptual model}
(Figure~\ref{fig:loop}) depicts the AI-enabled engineering process as a closed
socio-technical loop: human-specified \emph{intent} and \emph{context} drive agentic
\emph{generation}; output passes through \emph{verification and validation} and
\emph{integration}; and a \emph{governance} layer---with a persistent human-in-the-loop
checkpoint---feeds learning back into intent and context. The model makes explicit
that automation is encircled, not unsupervised: trust is produced by the loop, not
assumed of the generator~\cite{roychoudhury2025trust,kohl2026abundant,sun2026embedded}.

\subsection{Testable Propositions}
The framework yields falsifiable propositions (Table~\ref{tab:props}) intended to
guide future empirical work---through controlled experiments, mining of agentic
contribution data (e.g., AIDev-style corpora~\cite{li2025aiteammates}), longitudinal
field studies, and Delphi panels~\cite{betz2025responsibility}.

\begin{table*}[t]
\centering
\caption{Testable propositions derived from the framework.}
\label{tab:props}
\footnotesize
\begin{tabularx}{\textwidth}{c Y L{4.5cm}}
\toprule
\textbf{ID} & \textbf{Proposition} & \textbf{Suggested test}\\
\midrule
P1 & As agentic autonomy increases, the share of engineering effort devoted to verification and integration rises relative to authorship. & Time-allocation studies; effort logs across teams at differing autonomy levels.\\
P2 & Net throughput from agentic coding is gated by verification capacity, not generation speed. & Throughput vs.\ review-capacity regression on agentic-contribution data.\\
P3 & Teams with explicit orchestration and governance practices achieve higher acceptance/merge rates for agentic contributions. & Comparative mining of merge outcomes by governance maturity.\\
P4 & Provider-aligned agent--reviewer loops raise short-term velocity but increase latent quality/security risk. & Longitudinal defect/security analysis of closed vs.\ heterogeneous review loops.\\
P5 & Engineers and teams scoring higher on cognitive/governance competencies outperform those selected on coding proficiency alone in AI-enabled settings. & Competency assessment correlated with delivery and quality outcomes.\\
P6 & Curricula integrating systems thinking, V\&V, and AI governance yield graduates more effective in AI-enabled organizations. & Controlled curriculum comparison; employer/graduate longitudinal outcomes.\\
P7 & Overreliance on AI without foundational depth degrades diagnostic capability, raising long-term maintenance cost. & Quasi-experiments on debugging performance vs.\ AI-reliance intensity.\\
P8 & In safety-critical domains, increasing autonomy beyond human-oversight checkpoints correlates with unacceptable assurance gaps. & Domain case studies in embedded/safety-critical settings.\\
P9 & The economic value attributed to engineers increasingly derives from intent specification and judgment rather than code volume. & Compensation/role analyses; value-attribution surveys over time.\\
\bottomrule
\end{tabularx}
\end{table*}

\section{Threats to Validity}
\label{sec:threats}
We discuss validity using the standard categories, adapted for an interpretive
synthesis rather than a controlled empirical study.

\textbf{Construct validity.} Key constructs (e.g., ``agentic,'' ``orchestration,''
``competency'') are used variably across sources, and terminological drift between
research communities compounds the difficulty. We mitigated this by defining
constructs explicitly, adopting consistent terminology throughout, and grounding each
framework element in corpus evidence; nonetheless, residual ambiguity remains and our
operationalizations should be regarded as provisional and open to refinement as the
vocabulary of the field stabilizes.

\textbf{Internal validity.} As a synthesis, we infer mechanisms rather than establish
causation. Empirical anchors such as AIDev are observational and may reflect early
adoption dynamics, novelty effects, or selection among enthusiastic adopters rather
than steady-state behavior~\cite{li2025aiteammates}. Associations---for instance,
between governance maturity and merge rates---are deliberately framed as
\emph{propositions} precisely because they are not yet causally established and await
controlled or longitudinal confirmation.

\textbf{External validity.} The corpus skews toward open-source evidence and toward
research-active organizations; proprietary, regulated, and resource-constrained
settings may differ substantially in tooling, incentives, and review culture.
Safety-critical domains, in particular, exhibit distinct constraints and assurance
obligations~\cite{sun2026embedded}. Generalization should therefore be cautious,
domain-aware, and accompanied by local validation before transfer.

\textbf{Conclusion validity and recency.} The field is moving quickly, and part of
the corpus consists of archival manuscripts that, while authored by identifiable
researchers and verified against primary records, have varying degrees of peer review.
We treated such sources as evidence of direction rather than settled fact, and
triangulated them with peer-reviewed roadmaps and foundational theory. Findings tied
to specific tools or vendors may date rapidly; the framework is intentionally
tool-agnostic so as to remain robust to such churn.

\textbf{Researcher bias.} Interpretive synthesis is inevitably shaped by analyst
judgment and prior commitments. We reduced this risk through transparent inclusion
criteria, explicit traceability from evidence to claims, and deliberate preservation
of disagreement among sources, but we acknowledge that alternative, equally valid
syntheses of the same corpus are possible and would be a welcome basis for further
scrutiny.

\section{Future Research Directions}
\label{sec:future}
Several directions follow directly from our analysis. First, the propositions in
Table~\ref{tab:props} invite empirical validation through controlled experiments,
large-scale mining of agentic-contribution data, and longitudinal field
studies~\cite{li2025aiteammates}. Second, the community needs \emph{verification-aware
productivity metrics} that extend existing multidimensional frameworks to account for
trust, validation cost, and the effort of reviewing machine-generated
work~\cite{forsgren2021space}. Third, research should develop and rigorously evaluate
\emph{orchestration and review-triage methodologies} for high-throughput agentic
workflows, including structured rationales and test plans generated by the agents
themselves~\cite{li2025aiteammates}. Fourth, \emph{governance and assurance
models}---specifying when agents may act autonomously, how their contributions are
validated, and how accountability is assigned---require formalization and empirical
evaluation~\cite{sun2026embedded,roychoudhury2025trust}. Fifth, \emph{deskilling and
learning dynamics} merit longitudinal study to understand how foundational judgment can
be preserved amid pervasive automation~\cite{abrahao2025humans,mastropaolo2024riseandfall}.
Sixth, the FSE-CF itself should be validated and refined through expert Delphi panels,
competency assessment instruments, and curriculum
experiments~\cite{betz2025responsibility,goth2025rse,marlowe2026curriculum}. Pursued
together, these directions would convert the framework's conceptual claims into
testable, evidence-based guidance for practice and education.
\section{Conclusion}
\label{sec:conclusion}
Generative AI and Agentic AI are transforming software engineering from a primarily
code-producing activity into a more strategic, supervisory, and systems-oriented
discipline. The evidence indicates that work is migrating away from routine production
toward higher-order responsibilities: specifying intent, orchestrating agents,
verifying and integrating output, and governing risk and accountability. This
migration is already visible empirically---agents contribute at scale, yet their
output is accepted less readily and structurally simpler than that of humans,
revealing that the binding constraint has shifted from production to trust. We
characterized three coexisting paradigms, mapped activities being automated,
augmented, and newly emerging, and projected plausible role trajectories. Our central
constructive contribution is the Future Software Engineering Competency Framework,
which organizes required capabilities into technical, cognitive, socio-technical,
governance, and organizational categories, and which---through an accompanying
competency matrix and transformation model---makes explicit that the competencies
least substitutable by AI and most critical to success are cognitive and governance
competencies. As code becomes abundant, the durable value of the software engineer
resides not in the volume of code produced but in disciplined intent specification,
critical judgment, and accountable oversight of intelligent systems. Organizations,
universities, and policymakers that internalize this shift---and that build the
foundations and the new competencies in tandem---will be best positioned to prepare
software engineers to remain effective in AI-enabled development environments.


\bibliographystyle{unsrt}  
\bibliography{refs}  

@article{li2025aiteammates,
  author  = {Li, Hao and Zhang, Haoxiang and Hassan, Ahmed E.},
  title   = {The Rise of {AI} Teammates in Software Engineering ({SE}) 3.0: How Autonomous Coding Agents Are Reshaping Software Engineering},
  journal = {arXiv preprint arXiv:2507.15003},
  year    = {2025},
  note    = {AIDev dataset available at \url{https://github.com/SAILResearch/AI_Teammates_in_SE3}},
  url     = {https://arxiv.org/abs/2507.15003}
}

@article{abrahao2025humans,
  author  = {Abrah{\~a}o, Silvia and Grundy, John and Pezz{\`e}, Mauro and Storey, Margaret-Anne and Tamburri, Damian A.},
  title   = {Software Engineering by and for Humans in an {AI} Era},
  journal = {ACM Transactions on Software Engineering and Methodology},
  volume  = {34},
  number  = {5},
  pages   = {1--46},
  year    = {2025},
  articleno = {129},
  doi     = {10.1145/3715111}
}

@article{betz2025responsibility,
  author  = {Betz, Stefanie and Penzenstadler, Birgit},
  title   = {With Great Power Comes Great Responsibility: The Role of Software Engineers},
  journal = {ACM Transactions on Software Engineering and Methodology},
  volume  = {34},
  number  = {5},
  pages   = {1--21},
  year    = {2025},
  articleno = {136},
  doi     = {10.1145/3715112}
}

@article{goth2025rse,
  author  = {Goth, Florian and Alves, Renato and Braun, Matthias and Castro, Leyla Jael and others},
  title   = {Foundational Competencies and Responsibilities of a Research Software Engineer: Current State and Suggestions for Future Directions},
  journal = {F1000Research},
  volume  = {13},
  pages   = {1429},
  year    = {2025},
  doi     = {10.12688/f1000research.157778.2}
}

@article{roychoudhury2025trust,
  author  = {Roychoudhury, Abhik and P{\u{a}}s{\u{a}}reanu, Corina and Pradel, Michael and Ray, Baishakhi},
  title   = {Agentic {AI} Software Engineers: Programming with Trust},
  journal = {arXiv preprint arXiv:2502.13767},
  year    = {2025},
  url     = {https://arxiv.org/abs/2502.13767}
}

@article{ahmed2025journey,
  author  = {Ahmed, Iftekhar and Aleti, Aldeida and Cai, Haipeng and Chatzigeorgiou, Alexander and He, Pinjia and Hu, Xing and Pezz{\`e}, Mauro and Poshyvanyk, Denys and Xia, Xin},
  title   = {Artificial Intelligence for Software Engineering: The Journey So Far and the Road Ahead},
  journal = {ACM Transactions on Software Engineering and Methodology},
  volume  = {34},
  number  = {5},
  pages   = {1--27},
  year    = {2025},
  articleno = {119},
  doi     = {10.1145/3719006}
}

@article{acharya2025genai,
  author  = {Acharya, Vivek},
  title   = {Generative {AI} and the Transformation of Software Development Practices},
  journal = {arXiv preprint arXiv:2510.10819},
  year    = {2025},
  url     = {https://arxiv.org/abs/2510.10819}
}

@article{sun2026embedded,
  author  = {Sun, Simin and Staron, Miroslaw},
  title   = {Agentic Pipelines in Embedded Software Engineering: Emerging Practices and Challenges},
  journal = {arXiv preprint arXiv:2601.10220},
  year    = {2026},
  url     = {https://arxiv.org/abs/2601.10220}
}

@article{mastropaolo2024riseandfall,
  author  = {Mastropaolo, Antonio and Escobar-Vel{\'a}squez, Camilo and Linares-V{\'a}squez, Mario},
  title   = {The Rise and Fall(?) of Software Engineering},
  journal = {arXiv preprint arXiv:2406.10141},
  year    = {2024},
  url     = {https://arxiv.org/abs/2406.10141}
}

@article{marlowe2026curriculum,
  author  = {Marlowe, Thomas J. and Ku, Cyril S. and Laracy, Joseph R. and Kirova, Vassilka D. and Herbert, Katherine G.},
  title   = {A Systemic View of a Software Engineering Education Curriculum: Requirements and Guidelines in the Era of Generative {AI}},
  journal = {Journal of Integrated Design and Process Science},
  year    = {2026},
  doi     = {10.1177/10920617251405471}
}

@inproceedings{bruhin2024genai,
  author    = {Bruhin, Olivia and Ebel, Philipp and Mueller, Leon and Li, Mahei Manhai},
  title     = {{GenAI} and Software Engineering: Strategies for Shaping the Core of Tomorrow's Software Engineering Practice},
  booktitle = {Proceedings of the International Conference on Information Systems (ICIS)},
  year      = {2024},
  publisher = {Association for Information Systems}
}

@article{alenezi2025innovations,
  author  = {Alenezi, Mamdouh and Akour, Mohammed},
  title   = {{AI}-Driven Innovations in Software Engineering: A Review of Current Practices and Future Directions},
  journal = {Applied Sciences},
  volume  = {15},
  number  = {3},
  pages   = {1344},
  year    = {2025},
  doi     = {10.3390/app15031344}
}

@inproceedings{kohl2026abundant,
  author    = {Kohl, Karina and Carro, Luigi},
  title     = {When Code Becomes Abundant: Redefining Software Engineering Around Orchestration and Verification},
  booktitle = {Proceedings of the 2026 IEEE/ACM 48th International Conference on Software Engineering: Future of Software Engineering (ICSE-FoSE)},
  year      = {2026},
  doi       = {10.1145/3793657.3793884}
}

@book{brooks1975mythical,
  author    = {Brooks, Frederick P.},
  title     = {The Mythical Man-Month: Essays on Software Engineering},
  publisher = {Addison-Wesley},
  year      = {1975}
}

@article{naur1985theory,
  author  = {Naur, Peter},
  title   = {Programming as Theory Building},
  journal = {Microprocessing and Microprogramming},
  volume  = {15},
  number  = {5},
  pages   = {253--261},
  year    = {1985},
  doi     = {10.1016/0165-6074(85)90032-8}
}

@inproceedings{vaswani2017attention,
  author    = {Vaswani, Ashish and Shazeer, Noam and Parmar, Niki and Uszkoreit, Jakob and Jones, Llion and Gomez, Aidan N. and Kaiser, Lukasz and Polosukhin, Illia},
  title     = {Attention Is All You Need},
  booktitle = {Advances in Neural Information Processing Systems (NeurIPS)},
  year      = {2017}
}

@article{chen2021codex,
  author  = {Chen, Mark and Tworek, Jerry and Jun, Heewoo and Yuan, Qiming and others},
  title   = {Evaluating Large Language Models Trained on Code},
  journal = {arXiv preprint arXiv:2107.03374},
  year    = {2021},
  url     = {https://arxiv.org/abs/2107.03374}
}

@inproceedings{jimenez2024swebench,
  author    = {Jimenez, Carlos E. and Yang, John and Wettig, Alexander and Yao, Shunyu and Pei, Kexin and Press, Ofir and Narasimhan, Karthik},
  title     = {{SWE}-bench: Can Language Models Resolve Real-World {GitHub} Issues?},
  booktitle = {Proceedings of the Twelfth International Conference on Learning Representations (ICLR)},
  year      = {2024},
  url       = {https://arxiv.org/abs/2310.06770}
}

@article{forsgren2021space,
  author  = {Forsgren, Nicole and Storey, Margaret-Anne and Maddila, Chandra and Zimmermann, Thomas and Houck, Brian and Butler, Jenna},
  title   = {The {SPACE} of Developer Productivity},
  journal = {ACM Queue},
  volume  = {19},
  number  = {1},
  pages   = {20--48},
  year    = {2021},
  doi     = {10.1145/3454122.3454124}
}

@inproceedings{amershi2019se4ml,
  author    = {Amershi, Saleema and Begel, Andrew and Bird, Christian and DeLine, Robert and Gall, Harald and Kamar, Ece and Nagappan, Nachiappan and Nushi, Besmira and Zimmermann, Thomas},
  title     = {Software Engineering for Machine Learning: A Case Study},
  booktitle = {Proceedings of the 41st International Conference on Software Engineering: Software Engineering in Practice (ICSE-SEIP)},
  pages     = {291--300},
  year      = {2019},
  doi       = {10.1109/ICSE-SEIP.2019.00042}
}

@inproceedings{sculley2015debt,
  author    = {Sculley, D. and Holt, Gary and Golovin, Daniel and Davydov, Eugene and Phillips, Todd and Ebner, Dietmar and Chaudhary, Vinay and Young, Michael and Crespo, Jean-Fran{\c{c}}ois and Dennison, Dan},
  title     = {Hidden Technical Debt in Machine Learning Systems},
  booktitle = {Advances in Neural Information Processing Systems (NeurIPS)},
  year      = {2015}
}

@book{sommerville2015software,
  title={Engineering software products},
  author={Sommerville, Ian},
  volume={355},
  year={2020},
  publisher={Pearson London, UK}
}

@inproceedings{hamza2024human,
  title={Human-ai collaboration in software engineering: Lessons learned from a hands-on workshop},
  author={Hamza, Muhammad and Siemon, Dominik and Akbar, Muhammad Azeem and Rahman, Tahsinur},
  booktitle={Proceedings of the 7th ACM/IEEE International Workshop on Software-intensive Business},
  pages={7--14},
  year={2024}
}

@inproceedings{treude2025developers,
  title={How developers interact with AI: A taxonomy of human-AI collaboration in software engineering},
  author={Treude, Christoph and Gerosa, Marco A},
  booktitle={2025 IEEE/ACM Second International Conference on AI Foundation Models and Software Engineering (Forge)},
  pages={236--240},
  year={2025},
  organization={IEEE}
}

@article{peng2023copilot,
  author  = {Peng, Sida and Kalliamvakou, Eirini and Cihon, Peter and Demirer, Mert},
  title   = {The Impact of {AI} on Developer Productivity: Evidence from {GitHub} {Copilot}},
  journal = {arXiv preprint arXiv:2302.06590},
  year    = {2023},
  doi     = {10.48550/arXiv.2302.06590}
}

@inproceedings{vaithilingam2022expectation,
  author    = {Vaithilingam, Priyan and Zhang, Tianyi and Glassman, Elena L.},
  title     = {Expectation vs.\ Experience: Evaluating the Usability of Code Generation Tools Powered by Large Language Models},
  booktitle = {Extended Abstracts of the 2022 CHI Conference on Human Factors in Computing Systems (CHI EA '22)},
  year      = {2022},
  publisher = {ACM},
  doi       = {10.1145/3491101.3519665}
}

@article{barke2023grounded,
  author    = {Barke, Shraddha and James, Michael B. and Polikarpova, Nadia},
  title     = {Grounded {Copilot}: How Programmers Interact with Code-Generating Models},
  journal   = {Proceedings of the ACM on Programming Languages},
  volume    = {7},
  number    = {OOPSLA1},
  pages     = {85--111},
  year      = {2023},
  publisher = {ACM},
  doi       = {10.1145/3586030}
}

@inproceedings{liang2024survey,
  author    = {Liang, Jenny T. and Yang, Chenyang and Myers, Brad A.},
  title     = {A Large-Scale Survey on the Usability of {AI} Programming Assistants: Successes and Challenges},
  booktitle = {Proceedings of the IEEE/ACM 46th International Conference on Software Engineering (ICSE '24)},
  year      = {2024},
  publisher = {ACM},
  doi       = {10.1145/3597503.3608128}
}

@article{alenezi2026rethinking,
  title={Rethinking Software Engineering for Agentic AI Systems},
  author={Alenezi, Mamdouh},
  journal={arXiv preprint arXiv:2604.10599},
  year={2026}
}
\end{document}